\documentstyle[12pt]{article}
\begin{document}

\begin{center}

{\Large {\bf Nuclear Polarizabilities and Logarithmic Sum Rules}}\\

\vspace*{0.80in}

J.L.\ Friar \\
Theoretical Division \\
Los Alamos National Laboratory \\
Los Alamos, NM  87545 \\

\vspace*{0.20in}
and
\vspace*{0.20in}

G.\ L.\ Payne\\
Department of Physics and Astronomy\\
University of Iowa\\
Iowa City, IA 52242\\

\end{center}

\vspace*{0.50in}

\begin{abstract}

\noindent The electric polarizability and logarithmic mean-excitation  
energy are calculated for the deuteron using techniques introduced in  
atomic physics.  These results are then used to improve limits on the atomic-%
deuterium frequency shift due to nuclear polarization in the  
unretarded dipole limit, as well as confirming previous results.\\

\end{abstract}

\pagebreak

The remarkable experiments \cite{1,2} currently being performed on  
the isotope shift in atomic hydrogen ($^2$H vs.\ $^1$H) are primarily 
determined by differences in the masses of the isotopes, but are 
significantly sensitive to nuclear structure. These measurements 
provide the most precise determination of the difference in sizes of these
isotopes. The most recent \cite{2}  
result for the d-p isotope shift in the 1S-2S level splitting is
\begin{equation}
\Delta \nu_{\rm D-H} = 670994334 (2) \; {\rm kHz}\, ,
\end{equation}
of which roughly 5000 kHz is attributable to the finite-size differences of 
the nuclei, while roughly 20 kHz is due to the electric polarizability of 
the deuteron. In other words, in addition to a weaker Coulomb potential 
arising  
from the nuclear charge distribution seen by the electron at very short  
(on the atomic scale) distances, that electron also ``distorts'' or polarizes
the nucleus, which enhances the binding.  Four numerical calculations of  
the effect of nuclear polarization on the isotope shift have been  
performed recently \cite{3,4,5,5x}, although the relevant leading-order 
analytic results for the n\underline{th} S-state have long been 
known \cite{6,7}:
\begin{equation}
\Delta E_{\rm pol} = -5 m_e \; \alpha \; | \phi_n (0) |^2 \; \alpha_E \left(  
\frac{19}{30} + \log \left( \frac{2 \bar{E}}{m_e}\right) \right)\, ,
\end{equation}
where \cite{8} $\alpha$ is the fine-structure constant, $m_e$ is the  
electron mass, $|\phi_n (0)|^2 = \mu^3 \alpha^3 / \pi n^3$ is  
the square of the wave function of the electron at the origin,  
$\alpha_E$ is the deuteron electric polarizability, $\mu$ is the $e-d$  
reduced mass, and we work in natural units $(\hbar = c = 1)$. Even though 
uncertainties in the polarization calculations are currently smaller
than the error quoted in Eq.\ (1), planned improvements\cite{2} in that 
accuracy warrant a strong effort to reduce the theoretical uncertainty
to a minimum.

The electric polarizability of a nucleus (or atom) is defined by \cite{9}
\begin{equation}
\alpha_E = \frac{2 \alpha}{3} \; \sum_{N \neq 0} \; \frac{| \langle N |  
\vec{D} | 0 \rangle |^2}{E_N - E_0} \, ,
\end{equation}
where $E_0$ is the energy of the ground-state $|0\rangle$, $E_N$ is the 
energy of the N${\underline{th}}$ excited state, and $\vec{D}$ is the  
electric-dipole operator, which effects the transition between those  
states.  The definition (3) can be rearranged into the form of a sum rule  
\cite{9,10}:
\begin{equation}
\alpha_E = \frac{1}{2 \pi^2} \int d \omega \frac{\sigma^{ud}_{\gamma}
(\omega)}{\omega^2} \, \equiv \frac{\sigma_{-2}}{2 \pi^2}\, ,
\end{equation}
where $\sigma^{ud}_{\gamma} (\omega)$ is the cross section for  
photoabsorption of unretarded-dipole (long-wavelength) photons by  
the nucleus.  Concomitantly, the logarithmic mean-excitation energy
in Eq.\ (2), $\bar{E}$, is defined by
\begin{equation}
\frac{2 \alpha}{3} \; \sum_{N \neq 0} \; \frac{ | \langle N | \vec{D} 
| 0 \rangle |^2}{E_N - E_0} \; \log \; [ (E_N - E_0)/m_e] \equiv \alpha_E \; 
\log (\bar{E}/m_e)\, ,
\end{equation}
and clearly corresponds to placing a factor of $\log (\omega)$ in the  
integrand in Eq.\ (4).  The august $\sigma_{-2}$ sum rule and its (less  
well-known \cite{11}) logarithmic relative $\sigma^\ell_{-2}$ have been  
used to evaluate $\alpha_E$ and $\bar{E}$ by explicitly constructing  
$\langle N | \vec{D} | 0 \rangle$ (or equivalently, $\sigma^{ud}_{\gamma}  
(\omega)$) and performing the integral numerically.  Results 
for $\alpha_E$ for  
many ``realistic'' potential models are known, although the two most  
recent calculations \cite{4,5} did not have any models in common.

In this work we will:  (1) calculate $\alpha_E$ for a set of models that  
subsumes most of those of Refs.\ \cite{4} and \cite{5} and includes  
several more; (2) calculate log $(\bar{E})$ for these models; (3) use  
novel (for nuclear, but not atomic, applications) numerical techniques  
for calculating both $\alpha_E$ and log ($\bar{E}$); (4) critically  
discuss the potential models and attempt to assign a subjective but  
credible uncertainty to the results.  In this way we will confirm the  
previous results, while shrinking the uncertainty associated with them.   
Our numerical techniques were first applied to atomic problems, but  
now find a comfortable home in nuclear physics.

The technique we use for calculating $\alpha_{E}$ was first used by  
Podolsky \cite{12} to treat dispersion in hydrogen atoms.  Definition  
(3) is fully equivalent to
\begin{equation}
\alpha_{E} = 2 \alpha \; \langle 0 | D_{z} | \Delta \Psi_z \rangle \, ,
\end{equation}
where
\begin{equation}
(H - E_0) \; | \Delta \Psi_z \rangle \; = \; D_z | 0 \rangle
\end{equation}
is solved subject to finite boundary conditions.  Note that  
$\vec{D}$ does not connect the ground state (the only bound state) of  
the deuteron to itself. Resolution of Eq.\ (7) into  
partial waves, incorporation of the nuclear (including tensor) force, and  
other minor (although  tedious and important)  details are contained in Ref.\  
\cite{9}, together with many analytic results for simple potentials. Our 
calculation will employ the usual nonrelativistic (classical) dipole operator.

The resulting procedure is only slightly more complex than solving for  
the deuteron ground state, and it is very stable.  We have calculated  
$\alpha_E$ for 14 different ``realistic'' nucleon-nucleon (NN)  
potential models.  Such models must contain OPEP (the one-pion-exchange 
potential), which dominates the binding of light nuclei, 
and they must fit the NN data reasonably well.  All of the models used in  
Refs.\ \cite{4} and \cite{5} are in this category, although the quality of the fits  
(of various potential parameters) to the data differs rather dramatically from 
case to case.
Most of those models could be characterized as ``first-generation''  
models.  Recently, the Nijmegen group and their collaborators  
\cite{13,14,15} have constructed ``second-generation'' models, which  
provide good- to very-good-quality fits to all NN data, even approaching  
$\chi^2$ per degree of freedom $\sim 1$. Such fits are sufficiently good that
they can be regarded as alternative phase-shift analyses. This does not 
necessarily imply that the  
underlying physics has a corresponding accuracy, since several of these  
models are purely phenomenological, except for the all-important and  
dominant OPEP that incorporates different pion masses in different  
(isospin) states.

We determine the logarithmic mean-excitation energy $\bar{E}$ (or  
logarithmic sum rule) using a trick developed for calculating  
various logarithmic mean-excitation energies in atoms\cite{16}, one of  
which is the Bethe logarithm.  If we add a parameter $\lambda \equiv  
\xi \cdot f$ to $(H - E_0)$ in Eq.\ (7), where $\xi$ is  
dimensionless and $f$ has the dimensions of energy, we can then  
define (and easily calculate)
\begin{equation}
\alpha_E (\xi) = \frac{2 \alpha}{3} \; \sum_{N \neq 0} \; \frac{| 
\langle N | \vec{D} | 0 \rangle |^2}{\xi f + E_N - E_0}\, ,
\end{equation}
where $\alpha_E (0)$ is the usual result.  The integral of $\alpha_E  
(\xi)$ from $0$ to $\Lambda$ (very large compared to $\bar{E}/f$)  
generates
\begin{equation}
\int^{\Lambda}_{0} d \xi \, \alpha_E (\xi) \propto - \sum_{N \neq 0}  
\; \frac{| \langle N | \vec{D} | 0 \rangle |^2}{f} \; \log \; 
[ (E_N - E_0)/\Lambda f ] \, ,
\end{equation}
which gives the desired logarithm.  A similar integration gives
\begin{equation}
\int^{\infty}_{\epsilon} \frac{d \xi}{\xi} \, \alpha_E (\xi) \; \propto 
\sum_{N \neq 0} \; \frac{| \langle N | \vec{D} | 0 \rangle |^2}
{E_N - E_0} \; \log \; [(E_N - E_0) / \epsilon f]\, .
\end{equation}
For numerical purposes, we split the  
integral in Eq.\ (10) into $\int^{1}_{\epsilon} + \int^{\infty}_{1}$,  
and the dimensionful scale parameter $f$ determines where the split  
occurs in energy units.  Rearranging slightly and changing variables to  
$1/\xi$ in the second integral, we achieve our final result
\begin{equation}
\alpha_E (0) \, \log \; (2 \bar{E}/m_e) = \int^{1}_{0} \; 
\frac{d \xi}{\xi} [ \alpha_E (\xi) - \alpha_E (0) + \alpha_E 
( 1 / \xi )] - \alpha_E (0)\, \log (m_e/2 f) \, .
\end{equation}
The integrand is finite everywhere.  Choosing $f \sim 3-5 \, | E_0 |$  
makes the integral converge to 5 significant figures with only a few  
$(\sim 6)$ Gauss quadrature points, and all results for $\bar{E}$ are 
independent of  
$f$ if the integrals are performed with sufficient accuracy.  Podolsky's  
method \cite{9,12} makes the calculation of $\alpha_E (\xi)$ as  
easy as that of $\alpha_E (0)$.  The method is very stable.

\begin{table}[htb]
\centering

\caption{Deuteron electric polarizabilities, $\alpha_{\rm E}$, in units of 
fm$^{3}$, logarithmic mean-excitation-energy ratios, $\log(2 \bar{E}/m_e)$, 
and deuteron 1S-2S polarization-energy shifts, $\nu_{\rm pol}$, in kHz.}

\hspace{0.25in}

\begin{tabular}{|l|ccc|} \hline

\multicolumn{1}{|c|}{\rule{0in}{3ex} Potential Model}&
\multicolumn{1}{c}{$\alpha_{\rm E} ({\rm fm}^{3})$}&
\multicolumn{1}{c}{$\log(2 \bar{E}/m_e)$}&
\multicolumn{1}{c|}{$\nu_{\rm pol} ({\rm kHz})$}\\[0.5ex] \hline \hline
\multicolumn{4}{|c|}{Second-Generation Potentials}\\ \hline

Reid Soft Core (93)\rule{0in}{2.5ex}    & 0.6345 & 2.9616 &  19.31  \\
Argonne V$_{18}$      & 0.6343 & 2.9625 &  19.31  \\
Nijmegen (loc-rel)    & 0.6334 & 2.9618 &  19.28  \\
Nijmegen (loc-nr)     & 0.6327 & 2.9624 &  19.26  \\
Nijmegen (nl-rel)     & 0.6328 & 2.9619 &  19.26  \\
Nijmegen (nl-nr)      & 0.6319 & 2.9625 &  19.24  \\
Nijmegen (full-rel)   & 0.6311 & 2.9615 &  19.21  \\ \hline

\multicolumn{4}{|c|}{First-Generation Potentials} \\ \hline
                                    
Reid Soft Core (68) \rule{0in}{2.5ex}     & 0.6237 & 2.9638 &  18.99  \\
Bonn (CS)                & 0.6336 & 2.9630 &  19.29  \\
Paris                    & 0.6352 & 2.9627 &  19.34  \\
de Tourreil-Rouben-Sprung& 0.6376 & 2.9623 &  19.41  \\
Argonne V$_{14}$         & 0.6419 & 2.9624 &  19.54  \\
Nijmegen (78)          & 0.6472 & 2.9612 &  19.70  \\
Super Soft Core (C)      & 0.6497 & 2.9617 &  19.77  \\ \hline

\end{tabular}

\end{table}

Table 1 presents our results for $\alpha_E$, $\log ( 2 \bar{E} / m_e)$, and  
$\Delta E_{\rm pol}$ separated into first-generation  
\cite{17,18,19,20,21,22,23} (listed in order of appearance in Table 1) 
and second-generation \cite{13,14,15} (potential) categories.  Note that 
there is much more spread in the first-generation results, reflecting 
indifferent fits to the NN data.  The spread in the second-generation 
results can be summarized by
\begin{equation}
\nu_{\rm pol} = 19.26 (6) \, {\rm kHz}\, ,
\end{equation}
and
\begin{equation}
\alpha_E = 0.6328(17)\, {\rm fm}^{3}\, .
\end{equation}
As noted below in Ref.\ \cite{4}, this is not a numerically complete result
for the sum of all polarizability corrections, since it incorporates only 
unretarded dipole approximation. Higher multipoles, retarded dipole 
contributions, seagulls, etc., have not been included here, and may
decrease this result by up to 1 kHz\cite{4}.

All of the appropriate results are quite close to those previously 
calculated \cite{4,5},
with the electric polarizabilities differing at most by 2 in the last quoted  
significant figure in those references.  Such small differences could be  
attributed to slightly different versions of the potentials (new  
potentials are often a matter of ``work in progress'').  Note also that  
the pairs of new Nijmegen \cite{14} local and nonlocal potentials  
(labeled ``loc'' and ``$nl$'' in Table 1) have versions with relativistic 
(``rel'') and nonrelativistic (``nr'') kinematics (corresponding to 
{\it identical} deuteron energies of $2 \sqrt{M^2 - \kappa_{\rm rel}^2} - 2 M$ 
or $ - \frac{\kappa_{\rm nr}^2}{M}$, respectively). The slightly smaller 
value of $\kappa_{\rm rel}$ in the (excellent) zero-range approximation
\cite{5,9}
accounts for those differences in the values of $\alpha_E$,  although  
this makes relatively little difference in $\nu_{\rm pol}$.  The ``full''  
Nijmegen potential \cite{14} has the same form in all partial-waves
and fits the NN data less well than the others.

The result (13) agrees very well with a prediction\cite{9} of $\alpha_E$ = 
0.632(3)\, fm$^{3}$ 
made many years ago, and this warrants further comment. One can perform
perturbation theory about the ``zero-range'' limit by turning off the
forces in p-waves, dropping the deuteron d-state, and replacing the 
(reduced) deuteron s-state wave function by its asymptotic form: 
$u(r)= A_S \exp (-\kappa r)$, where $A_S$ is the s-wave asymptotic 
normalization constant. With this {\it ansatz} we obtain\cite{5,9}
\begin{equation}
\alpha_E \cong \alpha_E^0 = \frac{\alpha \mu A_S^2 }{32 \kappa^5}\, ,
\end{equation}
where $\mu$ is here the n-p reduced mass, and $\log(2 \bar{E}/m_e)$ = 
2.9671. This remarkably simple formula 
overestimates the complete result by approximately 1\% . There is little
uncertainty in any of the quantities except for $A_S$, which was recently 
determined\cite{24} to be $A_S$ = 0.8845(8) fm$^{-1/2}$ in agreement with 
the value
used in Ref.\ \cite{9}, and which leads to $\alpha_E^0$ = 0.6378(12) 
fm$^{3}$. Moreover, the corrections, $\Delta \alpha_E$,  to $\alpha_E^0$ 
defined by $\alpha_E = \alpha_E^0 + \Delta \alpha_E$ can be determined 
from the potential models (see Refs.\ \cite{5} and \cite{9}) 
to be $\Delta \alpha_E \cong -0.0044(2)$ fm$^{3}$, which leads directly to
$\alpha_E$ = 0.6334(14) fm$^3$, which is consistent with Eq.\ (13). Note that 
no relativistic corrections have been incorporated and they are not likely to
be negligible on the scale of the uncertainty in Eq.\ (13).

Why do the ``second-generation'' potentials agree so well with the perturbation
theory estimates? The answer is that $A_S$ is determined by analyzing NN
scattering, and we stated earlier that the new potentials could be viewed as
alternative phase-shift analyses. That is, they fit the NN data quite well, and 
associated properties (such as $A_S$) should agree with other experimental
determinations. Thus, $\alpha_E$ is very well determined.

We summarize by noting that $\alpha_E$ and $\log (\bar{E})$ have been 
calculated for the deuteron by novel methods.  These calculations confirm 
previous results and add additional ones.  We strongly recommend that  
only second-generation potential results be used when assessing the  
reliability of $\alpha_E$ calculations.  Equation (12) gives our best  
estimate for the leading-order (unretarded-dipole or long-wavelength)  
approximation to the nuclear-polarizability correction given by Eq.\  
(2).\\

\noindent \underline{Acknowledgements} 

The work of JLF was performed under the auspices of the U.S.\  
Department of Energy, while that of GLP was supported in part by the  
U.S.\ Department of Energy.  One of us (JLF) would like to thank Don  
Sprung and Winfried Leidemann for helpful discussions of their  
results. We would also like to thank S.\ Karshenboim for pointing
out Ref.\ \cite{5x} to us. \\

\end{document}